# Confidential Encrypted Data Hiding and Retrieval Using QR Authentication System

Somdip Dey
Department of Computer Science
St. Xavier's College [Autonomous]
Kolkata, India
e-mail: somdipdey@ieee.org

Asoke Nath
Department of Computer Science
St. Xavier's College [Autonomous]
Kolkata, India
e-mail: asokejoy1@gmail.com

Shalabh Agarwal
Department of Computer Science
St. Xavier's College [Autonomous]
Kolkata, India
e-mail: shalabh@sxccal.edu

*Abstract*—Now, security and authenticity of data is a big challenge. To solve this problem, we propose an innovative method to authenticate the digital documents. In this paper, we propose a new method, where the marks obtained by a candidate will also be encoded in QR Code[TM] in encrypted form, so that if an intruder tries to change the marks in the mark sheet then he can not do that in the QR Code[TM], because the encryption key is unknown to him. In this method, we encrypt the mark sheet data using the TTJSA encryption algorithm. The encrypted marks are entered inside QR code and that QR code is also printed with the original data of the mark sheet. The marks can then be retrieved from the QR code and can be decrypted using TTJSA decryption algorithm and then it can be verified with marks already there in the mark sheet.

Keywords-communication technology; QR Code[TM]; encryption; decryption; multimedia; internet;

## I. INTRODUCTION

In this era of digital world, with the evolution of technology and un-ending growth in digital data, there is an essential need of optimization of online data and information present in the digital world. The most important issue in data is that it should be original and correct. The authenticity of data is the most challenging issue in management of data in the internet database. In the present study we mainly focus on authenticity of marks in a printed mark sheet. However the present method can be applied to any other legal documents also. We know every year billions of students pass and have passed from different schools, colleges and Universities all over the world. At the same time there is no scientific method to test the authenticity of data from any printed document. Mostly we depend on checking by human eye where human eye may not function always perfect. Moreover there is no second verification on human eye verification of document. Keeping this problem in mind, we have introduced a new digital mark-sheet system. In our new mark-sheet system, we will be embedding the data digitally in form of QR Code [9][11], which is itself encrypted, so that the marks obtained by the student can not be tampered, and the data embedded in the mark-sheet can be only decrypted and read from our decryption program. In this way, we do not have to increase our digital space or add new servers to our already existing system just because to save more marks record of students.

QR Code[TM] [9][11] is a type of 2 dimensional matrix barcode, which gained popularity because of its large capacity to hold digital data and it can be integrated in any mobile devices. In our new mark-sheet system, we save the essential data of each student in the QR Code, like the student's name, roll number, registration number, semester and year of study, marks obtained in different subjects and grades secured. But, all the data saved and embedded in the QR Code, are encrypted, and then the QR Codes are printed in the mark-sheet of the student. So, in future if the student or any other person wants to see their marks digitally or wants to send their academic information to any University or Organization in digital format, then they can just scan the QR Code and decrypt the embedded information and send the authentic data.

## II. METHODS USED

We use TTJSA [1] encryption algorithm, which was designed by Nath et al. and is an amalgamation of three different cryptographic modules: generalized modified Vernam cipher [1], MSA [2] and NJJSA [3], for the encryption purpose of data in the QR Code. After encrypting the data, we embed the data in the QR Code using a set of different protocols and ultimately generate the encrypted QR Code. We discuss the procedure elaborately in the following sections.

*A. TTJSA for Encryption Purpose of the Embedded Data*

TTJSA [1] is a combined symmetric key cryptographic method, which is formed of generalized modified Vernam cipher, MSA and NJJSA symmetric key cryptographic methods. Brief study of the methods used in TTJSA algorithm is as follows:

*1) Modified Vernam Cipher*

In this step, we break the whole file into different small blocks (like in Block Cipher system []), where each block size should be less than or equal to 256 byes. Then we follow these steps:

Step1: Perform normal Vernam Cipher method with the block of randomized key i.e. each byte of blocks of the file + each byte of the blocks of randomized key.

Step 2: If the pointer reaches the end of each block then after performing Vernam Cipher method, pass the remainder of the addition of the last byte of the file block with the last byte of the key to the next file block and add the remainder with the first byte of the that file block. (This mechanism is called feedback mechanism)

Step 3: Perform Step 1 and Step 2 until the whole file is encrypted and repeat this step for random number of times.

After performing the aforementioned steps, we again merge the blocks of the encrypted file and thus we get the final encrypted result of this modified Vernam Cipher method.

*2) NJJSAA Algorithm*

The encryption number (=secure) and randomization number (=times) is calculated according to the method mentioned in MSA algorithm [2].

Step 1: Read 32 bytes at a time from the input file.
Step 2: Convert 32 bytes into 256 bits and store in some 1-dimensional array.
Step 3: Choose the first bit from the bit stream and also the corresponding number(n) from the key matrix. Interchange the 1st bit and the n-th bit of the bit stream.
Step 4: Repeat step-3 for 2nd bit, 3rd bit...256-th bit of the bit stream
Step 5: Perform right shift by one bit.
Step 6: Perform bit(1) XOR bit(2), bit(3) XOR bit(4),...,bit(255) XOR bit(256)

Step 7: Repeat Step 5 with 2 bit right, 3 bit right,...,n bit right shift followed by Step 6 after each completion of right bit shift.

*3) MSA Encryption and Decryption Algorithm*

Nath et al. [2] proposed a symmetric key method where they have used a random key generator for generating the initial key and that key is used for encrypting the given source file. MSA method is basically a substitution method where we take 2 characters from any input file and then search the corresponding characters from the random key matrix and store the encrypted data in another file. MSA method provides us multiple encryptions and multiple decryptions. The key matrix (16x16) is formed from all characters (ASCII code 0 to 255) in a random order.

The randomization of key matrix is done using the following function calls:
Step-1: call Function cycling()
Step-2: call Function upshift()
Step-3: call Function downshift()
Step-4: call Function leftshift()
Step-5: call Function rightshift()

How the above functions will work have been discussed in detail by Nath et al[1]. The idea of these functions is to make elements in a square matrix in a random order so that no one can predict what will be the nearest neighbour of a particular element in that matrix. This method is basically modified Playfair method. In Playfair method one can only encrypt Alphabets but in MSA one can encrypt any character whose ASCII code from 0-255 and one can apply multiple encryption here which is not possible in normal Playfair method.

*B. Generation of QR Code*

To create a QR code [9][10][11] is we first create a string of data bits. This string includes the characters of the original message (encrypted message in this case) that you are encoding, as well as some information bits that will tell a QR decoder what type of QR Code it is.

After generating the aforementioned string of bits, we use it to generate the error correction code words for the QR Code. QR Codes use Reed-Solomon Error Correction technique [10][12] Please note that in coding theory, Reed-Solomon codes (RS codes) are non-binary cyclic error correction codes invented by Irving S. Reed and Gustave Solomon.

After the generation of bit-string and error correction code words, the resultant data is used to generate eight different QR Codes, each of which uses a different mask pattern. A mask pattern controls and changes the pixels to light or dark ones, according to a particular formula. The eight mask pattern formulas are defined in the QR Code specification, which is referred at the time of mask generation needed for the QR Code generation. Each of the eight QR codes is then given a penalty score that is based on rules defined in the QR specification. The purpose of this step is to make sure that the QR code doesn't contain patterns that might be difficult for a QR decoder to read, like large blocks of same-colored pixels, for example. After determining the best mask pattern, the QR Code, which uses the best mask pattern, is generated and shown as an output.

If the size of the encrypted message becomes more than 1,264 characters then the characters appearing after 1,264 characters are used separately to generate another QR Code and the above mentioned process is repeated until and unless the total encrypted message is converted to QR Code(s).

The method is discussed in details below:
The Encrypted file, which is created using the method TTJSA is now treated as the input file and the string is extracted from the file to generate the QR Code.
Step 1: call function file_read(output_file)
Step 2: call function generateQRCode( str[] )
Step 3: call function delete_file(output_file)

1) Algorithm for generateQRCode() :

 a) Step1:
**Mode Indicator**
e.g.: Numeric Mode: 0001, Alpha Numeric: 0010, Let us choose 0010 for Alphanumeric
**Character Count**
e.g.: Numeric: 10bit long, Alphanumeric: 8bit long (for version 1-9) Let's encode 8 in 8bit long binary representation 0010   000001000
**Encode Data**
Numeric Mode: Data delimited by 3digit Alphanumeric Mode: Data delimited by 2digit e.g.: Let's take ABCDE123"AB": 45*10+11=461   "CD":45*12+13=553 "E1": 45*14+1=631 "23": 45*2+3=93 Codeword for A=10, B=11, C=12 etc. Now the value encoded in 11bit long binary representation. 0010   000001000   00111001101 01000101001 01001110111 00001011101

**Termination**
Add 0000 at the end to terminate 0010 000001000 00111001101 01000101001 01001110111 00001011101 **0000**
**Encode to Code Word**
Result data are delimited by 8bit 00100000 01000001 11001101   01000101   00101001   11011100   00101110 10000 If last data is less than 8bit, pad it with 0 00100000 01000001   11001101   01000101   00101001   11011100 00101110 10000**000**. We alternatively put '11101100' and '00010001' until full capacity of the following version 00100000   01000001   11001101   01000101   00101001 11011100 00101110 10000000 **11101100**
Decimal Representation: 32 65 205 69 41 220 46 128 236

 b) Step 2:
**Reed Solomon Error correcting Code** [10][12] is used in QR Code
e.g.: For example data, count of error correcting
code word is 17 $g(x)=x^{17}+\alpha^{43}x^{16}+\alpha^{139}x^{15}+\alpha^{206}x^{14}+\alpha^{78}x^{13}$ $+\alpha^{43}x^{12}+\alpha^{239}x^{11}+\alpha^{123}x^{10}+\alpha^{206}x^9+\alpha^{214}x^8+\alpha^{147}x^7+\alpha^{24}x^6$ $+\alpha^{99}x^5+\alpha^{150}x^4+\alpha^{39}x^3+\alpha^{243}x^2+\alpha^{163}x+\alpha^{136}$
Now polynomial f(x) which coefficients are data code words is divided by g(x)
$f(x)=32x^{25}+65x^{24}+205x^{23}+69x^{22}+41x^{21}+220x^{20}+46x^{19}+128x^{18}+236x^{17}$ ----(i)
divided  by g(x)
$g(x)*(\alpha^5)*x^8$
$=\alpha^5*x^{25}+\alpha^5*\alpha^{43}*x^{24}+\alpha^5*\alpha^{139}*x^{23}+\alpha^5*\alpha^{206}*x^{22}+\alpha^5*\alpha^{78}*x^{21}$.....
$=\alpha^5*x^{25}+\alpha^{48}*x^{24}+\alpha^{144}*x^{23}+\alpha^{211}*x^{22}+\alpha^{83}*x^{21}$..... $=32x^{25}+70x^{24}+168x^{23}+178x^{22}+187x^{21}$......-----(ii)
Calculate Exclusive logical Sum (i) and (ii)
$f(x)'=7x^{24}+101x^{23}+247x^{22}+146x^{21}$.....
We repeat same logic until this divide calculation is over.
Finally we get R(x).
$R(x)=42x^{16}+159x^{15}+74x^{14}+221x^{13}+244x^{12}+169x^{11}+239x^{10}+150x^9+138x^8+70x^7+237x^6+85x^5+224x^4+96x^3+74x^2+219x+61$
So we get    32 65 205 69 41 220 46 128 236 **42 159 74 221 244 169 239 150 138 70 237 85 224 96 74 219 61**

 c) Step 3:
**Data Allocation**
Step-1: Start module is lower right corner.
Step-2: We think 2 modules width. If we are in right module of 2 modules width then, If left module is blank (not fixed pattern or version information etc), we move left module and put data. If left module is not blank, we move in direction, which is kept, and put data.
If we are in left module then, we check that blank module is in direction which is kept. If blank module is, we put data in right module in priority to left module of 2 modules width.
If blank module is not, we move to a left module, and put data there. Then we turn direction which is kept.
Step-3: Direction of movement is upper or lower. First its upper then its lower.
e.g.: If we have data "89ABCDEF GHIJKLMN" and we put it in 6*4 matrix.

```
D C B A
F E 9 8
H G 7 6
J I 5 4
L K 3 2
N M 1 0
```

 d) Step 4:
**Mask Pattern**
We select from 8 mask pattern
000: (i+j) mod 2 = 0
001: i mod 2 = 0
010: j mod 3 = 0
011: (i+j) mod 3 = 0
100: ((i div 2) + (j div 3)) mod 2 = 0
101: (ij) mod 2 + (ij) mod 3 = 0
110: ((ij) mod 2 + (ij) mod 3) mod 2 = 0
111: ((ij) mod 3 + (i+j) mod 2) mod 2 = 0
"mod" means remainder calculation, "div" means integer divide.

 e) Step5:
**Format Information**
It includes error correcting level and mast pattern indicator in 15bit long.

First two bit is error correcting level
01: L Error correcting level
00: M Error correcting level
11: Q Error correcting level
10: H Error correcting level
Next three bit is mask pattern indicator and next 10bit we put error correcting data which is Bose-Chaudhuri-Hocquenghem (BCH).

  *f) Step 6:*
**Generate QR Code Image**
Library Class is used to generate the image.

### ALGORITHM FOR DECODE QRCODE()

We here follow the reverse process of the above *generateQRCode()* Algorithm to detect the QR Code Image using Library Class [11] and perform error correction using Reed-Solomon technique and get back the encrypted message.

### III. RESULTS AND DISCUSSION

We choose a student of anonymous name and produce the demonstration of the new mark-sheet system of that student in the following figures.

**RESULT 1:**

Fig 1: Student XYZ's Information in form of Encrypted QR Code

Decryption of Data          From QR Code

Fig 2: Decrypted Student XYZ's Information from QR Code

We have also given a demonstration of our marks-sheet in the following figure:

Fig 3: An Actual Result having the Digital Data in encrypted QR Code

**RESULT 2:**

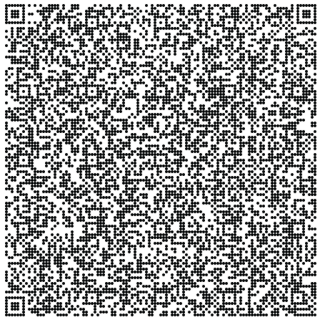

Fig 4: Student DEF's Information in form of Encrypted QR Code

Decryption of Data From QR Code

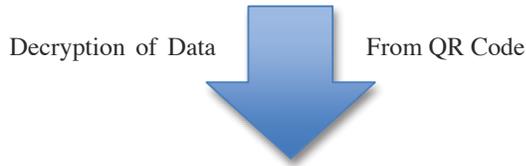

ABC College (Autonomous)
Affiliated to University of XXXXX

COMPUTER Sc. (HONS.) 1st. SEMESTER EXAMINATION, YEAR 20XX

The following is the statement of marks obtained by DEF
Roll : 0-00-00-0002
Regd No. : A00-0000-0000-02
At the aforesaid Examination held in NOV – DEC 20XX

SUBJECT
CMSA3101  : 70 OUT OF 100
CMSA3151  : 45 OUT OF 50
MBNG1101  : 40 OUT OF 50
MTMG2101  : 50 OUT OF 75
PHSG2101  : 35 OUT OF 50

1ST CLASS : 60%
2nd CLASS : 40%

Fig 5: Decrypted Student DEF's Information from QR Code

We have also given a demonstration of our marks-sheet in the following figure:

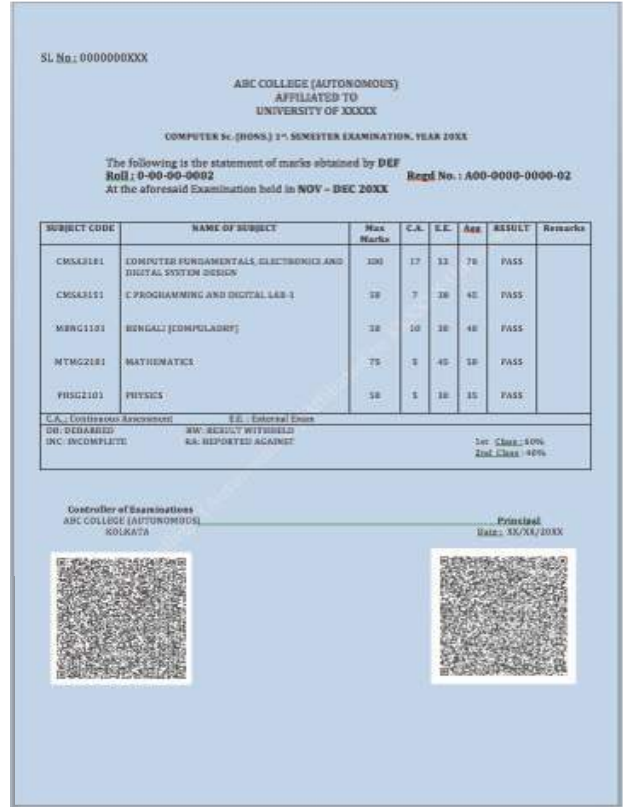

Fig 6: An Actual Result having the Digital Data in encrypted QR Code

## IV. SECURITY ANALYSIS

We chose to tamper the data of student XYZ and intentionally changed marks obtained by the student XYZ in only one subject. Then, we encrypted the data of the tampered mark-sheet. After that, we ran frequency analysis of both the encrypted data, and the frequency analysis of both the encrypted data were totally different and no pattern was found among them.

ABC College (Autonomous)
Affiliated to University of XXXXX

COMPUTER Sc. (HONS.) 1st. SEMESTER EXAMNATION, YEAR 20XX

The following is the statement of marks obtained by XYZ
Roll : 0-00-00-0001
Regd No. : A00-0000-0000-01
At the aforesaid Examination held in NOV – DEC 20XX

SUBJECT
CMSA3101  : 66 OUT OF 100
CMSA3151  : 43 OUT OF 50
MBNG1101  : 30 OUT OF 50
MTMG2101  : 50 OUT OF 75
PHSG2101  : 28 OUT OF 50

1ST CLASS : 60%
2nd CLASS : 40%

Fig 7: Original Data of Student XYZ

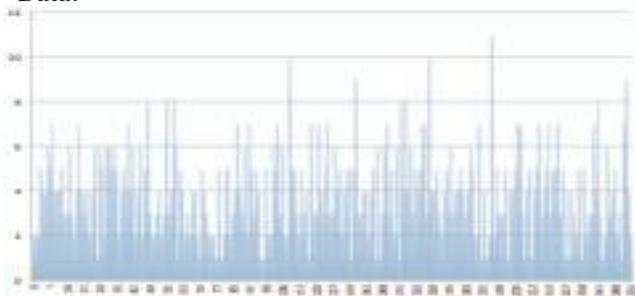
Fig 8: Tampered Data of Student XYZ

**Frequency Analysis of Original Data Vs Tampered Data:**

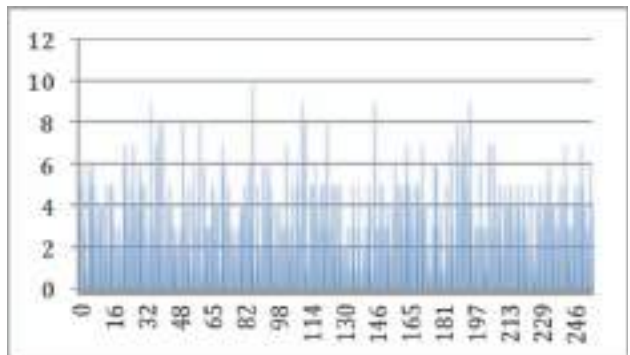
Fig 9: Frequency Analysis of Encrypted Original File

Fig 10: Frequency Analysis of Encrypted Tampered Data

Thus, from the frequency analysis (spectral analysis) it is evident that if the data is tampered then, the encrypted data of the tampered file will be very different from the encrypted data of the original one. And by comparing the frequency analysis of the two encrypted data, it can be verified whether the data is authentic (original) or not.

## V. CONCLUSION AND FUTURE SCOPE

In the present work we have mainly focus on confidential encrypted data hiding in QR Code. As we know that data embedding and retrieval from QR-code is very simple issue. Simply a smart phone running on Android or iOS or any other new generation of mobile OS, can be used to extract the encrypted data from embedded QR-code and finally that data to be decrypted using the TTJSA decryption algorithm.

ACKNOWLEDGMENT

Somdip Dey (SD) expresses his gratitude to all his fellow students and faculty members of the Computer Science Department of St. Xavier's College [Autonomous], Kolkata, India, for their support and enthusiasm. AN is grateful to Dr. Fr. Felix Raj, Principal St. Xavier's College, Kolkata for giving opportunity to work in the field of data hiding and retrieval.